\documentclass[11pt]{article}
    \usepackage[dvips]{graphicx}
    \usepackage{amsmath}
    \usepackage{amsxtra}
    \usepackage{txfonts}
    \usepackage{pxfonts}

    \usepackage[OT2,T1]{fontenc}

    \numberwithin{equation}{section}
\begin{document}

    \title{Classical Kinematics:\\[3pt]
    {\Large Derivation and New Interpretation\\
     of the Lorentz Transformations\\[-3pt]
     and Einstein's Theorem of Velocity Addition}}

   \author{Vladimir T. Granik\thanks {University of California, CEE Department (Ret.)\;\!, Berkeley, CA 94720, USA. Email: \mbox{vtgranik@comcast.net.}}\hspace{-28pt} \and \!and Alex Granik\thanks{University of the Pacific, Department of Physics, Stockton, CA 95211, USA. Email: agranik@pacific.edu.}}

    \date{}

    \maketitle

    \vspace{-16pt}

    \noindent

   \begin{quote}\small{It is traditionally believed that the Lorentz transformations (LT) and Einstein's theorem of velocity addition (ETVA), underlying special relativity, cannot be obtained from non-relativistic (classical) mechanics. In the present paper it is shown, however, that both the LT and the ETVA are derivable within the framework of classical kinematics if the speeds of material points are bounded above by a certain universal limit $c_+$ which can coincide with the speed of light $c$ in a vacuum.}

    \vspace {6pt}

    \end{quote}

        \thispagestyle{empty}
\normalsize
                                \section{\Large Introduction}
    \noindent The Lorentz transformations and Einstein's theorem of velocity addition, which form the basis of special relativity [1--3], are    traditionally considered not derivable from classical mechanics [cf. 4--7]. However, in the present paper it is shown for the first time that the above basic relations can be obtained within the framework of classical kinematics provided the speeds of material points are bounded above by some universal limit $c_+$.

    To begin with, consider three consecutive material points \textit{A}, \textit{B} and \textit{D} on the \textit{X}-axis of an inertial Cartesian frame of reference S. Point \textit{A} is the origin of the frame S, whereas point \textit{B} is the origin of a second inertial frame S'  moving with respect to S along their common axis \textit{X} at a constant translational speed $v$\footnote{\,In this case, the first frame S is conventionally regarded as the system of absolute coordinates (also called the fixed system), and the second frame S' is called the system of relative coordinates.}. Also, point \textit{D} travels along the positive direction of the \textit{X}-axis with respect to points \textit{B} and \textit{A}. This means that point \textit{D} executes relative motion in the system S' and absolute (compound) motion in the system S.

        At the moment $t' > 0$ of absolute time $\tau$ used in classical kinematics, points \textit{A}, \textit{B} and \textit{D} take positions \textit{A'}, \textit{B'} and \textit{D'}, so that the displacements $A'B'$, $B'D'$ and $A'D'$ are respectively $vt',\, x'(t')$ and $x(t')$. In accordance then with classical kinematics, $x(t') = vt' + x'(t')$. At some later moment $t > t'$ of absolute time $\tau$, the above displacements become $AB = vt, BD = x'(t)$ and $AD = x(t)$. Likewise, $x(t) = vt + x'(t)$. Since point $D$ moves away from point $A$, the displacement $AD > A'D'$ or $x(t) > x(t')$. We thus have the following chain of kinematic relations:
                \begin{equation}
                        x(t) = x'(t) + vt > x(t') = x'(t') + vt' \quad(t > t').
                \end{equation}

    According to (1.1), $x(t) > x'(t') +\, vt'$ or alternatively
                \begin{equation}
                        x(t) = {\lambda_1}(x'(t') + vt'),
                \end{equation}
    where $\lambda_1 > 1$ is some finite factor to be determined below in Section 3.

    From (1.1) it also follows that $x'(t)\, +\, vt > x'(t')\, +\, vt'$ or $x'(t') < x'(t) + v(t - t')$. Because $t > t'$, we have $x'(t') \neq x'(t)$ which in view of the classical identity $x'(t) = x(t) - vt$\, takes the form $x'(t') \neq x(t) - vt$ or
                \begin{equation}
                        x'(t') = \lambda_2(x(t) - vt),
                \end{equation}

    \noindent where $\lambda_2 > 0$ is a second finite factor to be determined in Section 3.

    As shown below, the classical kinematic relations (1.2) and (1.3) lead first to the theorem of velocity composition (Section 2) and then to the Lorentz transformations (Section 4) and Einstein's theorem of velocity addition (Section 5).
                                \section{\Large The theorem of velocity composition}
    In addition to classical kinematics, we assume that the speeds of any material point in the inertial frames S and S' are bounded above by some universal limit $c_+$. Let then $u'(t') > 0$ be the average relative speed of point \textit{D'} with respect to the origin \textit{B'} of system S' over the time interval $[\;\!0,t']$, whereas $u(t) > 0$\footnote{\,The cases of $u'(t') = 0$ or $u(t) = 0$ are ruled out because either makes it impossible to obtain the subsequent key relation (2.4) from (2.2) and (2.3).} be the average absolute speed of the same point $D$ with respect to the origin \textit{A} of system S over the time interval $[\;\!0,t\;\!]$. Consequently, these speeds are
                        \vspace{-2pt}

                \begin{equation}
                        u'(t') \equiv \frac{{x'(t')}}{t'} \leq c_+, \quad u(t) \equiv \frac{{x(t)}}{t}  \leq c_+ \quad (t > t' > 0).
                        \vspace{2pt}
                \end{equation}

    Now, in view of (2.1), Eqs. (1.2) and (1.3) take the form\footnote{\,For simplicity, the arguments $t$ and $t'$ of the functions $u(t)$ and $u(t')$ are dropped in (2.2), (2.3) and in some ensuing relations.}
                            \vspace{-1pt}
                \begin{gather}
                        ut = {\lambda_1}t'(u' + v),\\
                        u't' = {\lambda_2}t(u - v).
                \end{gather}
    The product $\lambda_1\lambda_2$ of (2.2) and (2.3), divided by $t\,\!t' \neq 0$, yields \vspace{1pt}
                \begin{equation}
                        \lambda_1\lambda_2 = \frac{{uu'}}{{(u - v)}{(u' + v)}} \quad (u > 0,\;\, u' >0,\;\, t\,\!t' \neq 0).
                        \vspace{4pt}
                \end{equation}
    From this relation we find (including the upper bound limitation $u \leq c_+$) \vspace{3pt}
                \begin{equation}
                        u = \frac{\lambda_1\lambda_2v(u' + v)}{\lambda_1\lambda_2(u' + v) - u'} \leq c_+,
                        \vspace{2pt}
                        \end{equation}
    whence $\lambda_1\lambda_2 v(u' + v) \leq c_+(\lambda_1\lambda_2(u' + v) - u')$. Solving this weak inequality for $u'$, we obtain
                \begin{equation}
                        u' \leq \frac{\lambda_1\lambda_2v(c_+ - v)}{c_+ + \lambda_1\lambda_2(v - c_+)}.
                        \vspace{4pt}
                \end{equation}
    The upper bound for the speed $u'$ given by the right-hand side of (2.6) must coincide with the upper bound $c_+$ for the same speed according to (2.1), i.e. $\lambda_1\lambda_2v(c_+ - v)/(c_+ + \lambda_1\lambda_2 (v - c_+)) = c_+$, whence it follows that
                \begin{equation}
                       \lambda_1\lambda_2 = \frac{c_+^{2}}{c_+^{2} - v^2}\,.
                \end{equation}
    Due to (2.7), the right-hand side of (2.4) becomes
                \begin{equation}
                        \frac{{uu'}}{{(u - v)}{(u' + v)}} = \frac{c_+^{2}}{c_+^{2} - v^2} \quad (u > 0,\;\, u' >0,\;\, t\,\!t' \neq 0). \vspace{1pt}
                \end{equation}

    Finally, Eq. (2.8) can be solved for $u$ and $u'$ to yield
                \begin{gather}
                        u(t) = \frac{u'(t') + v}{1 + u'(t')v/c_+^2},\\                         u'(t') = \frac{u(t) - v}{1 - u(t)v/c_+^2}.
                \end{gather}

    \noindent Formulae (2.9) and (2.10) represent the theorem of velocity composition in non-traditional classical kinematics in which the absolute $u(t)$ and the relative $u'(t')$ speeds depend on time and describe a non-uniform rectilinear motion of material point $D$ with respect to the inertial frames S and S' along their common axis $X$. Unlike traditional classical kinematics, the above speeds cannot be infinitely large because of being bounded above by the finite limit $c_+$. If, however, one assumes that $c_+ = \infty$, then (2.9) and (2.10) reduce to the classical theorem of velocity addition
                \begin{equation}
                        u = u' + v.
                \end{equation}

    According to (2.9), if at the moment $t'$ the relative speed $u'(t')$ reaches the upper bound $c_+$, i.e. $u'(t') = c_+$,\, then the absolute speed $u(t)$ reaches the same upper bound $c_+$, i.e. $u(t) = c_+$,\; at a later moment $t > t'$.

    On the other hand, as follows from (2.10), the condition $u(t) = c_+$\, at the moment $t$ necessitates the condition $u'(t') = c_+$\, at an earlier moment $t' < t$.

    Hence, in consequence of the theorem of velocity composition (2.9)--(2.10), the same upper bound $c_+$ can be reached by both the relative speed $u'$ and the absolute speed $u$ but only at different moments of time.
                    \section{\Large Determination of the factors $\lambda_1$ and $\lambda_2$}
    From (2.9) and (2.10) it follows respectively that
                            \vspace{-1.5pt}
                \begin{gather}
                        u' + v = u(1 + vu'/c_+^2),\\
                        u - v  = u'(1 - uv/c_+^2).
                \end{gather}
    Substitution of (3.1) into (2.2) and (3.2) into (2.3) gives
                \begin{gather}
                        ut = \lambda_1 t'u(1 + vu'/c_+^2),\\
                        u't' = \lambda_2 tu'(1 - vu/c_+^2).
                \end{gather}
    Canceling $u > 0$ from both sides of (3.3) and $u' > 0$ from both sides of (3.4), we have
                            \vspace{-8pt}
                \begin{gather}
                        t = \lambda_1 t'(1 + vu'/c_+^2),\\
                        t' = \lambda_2 t(1 - vu/c_+^2).
                \end{gather}
    In view of (2.1), Eqs. (3.5) and (3.6) take the form
                \begin{gather}
                        t = \lambda_1(t' + vx'(t')/c_+^2),\\
                        t' = \lambda_2(t - vx(t)/c_+^2).
                \end{gather}
    \indent Now introduce dimensionless speeds:
                \begin{gather}
                        \alpha(t) \equiv u(t)/c_+ \in (0, 1],\quad \text{i.e.}\quad 0 <\, \alpha(t) \leq 1,\\
                        \alpha'(t') \equiv u'(t')/c_+ \in (0, 1],\quad \text{i.e.}\quad 0 <\, \alpha'(t') \leq 1,\\
                        \beta_+ \equiv v/c_+ = \textrm{constant} \in (0, 1),\quad \text{i.e.}\quad 0 <\, \beta_+ < 1.
                \end{gather}
     Due to (3.9)--(3.11), Eqs. (3.5) and (3.6) become\footnote{\,For simplicity, the arguments $t$ and $t'$ of the functions $\alpha(t)$ and $\alpha'(t')$ are dropped in  (3.12), (3.13) and in\, all ensuing relations.}
                \begin{gather}
                        t = \lambda_1 t'(1 + \alpha'\beta_+),\\
                        t' = \lambda_2 t(1 - \alpha\beta_+).
                \end{gather}
    Also recall that $t > t'$ and hence $t' < t$. Making use of these inequalities in (3.12) and (3.13), respectively, we obtain after a little algebra
                \begin{gather}
                        \lambda_1 > 1/(1 + \alpha'\beta_+),\\
                        \lambda_2 < 1/(1 - \alpha\beta_+).
                \end{gather}

    Next, assume that $\lambda_2 = \eta\lambda_1$ so that (3.14) and (3.15) can be written as one triple inequality
                \begin{equation}
                        1/(1 + \alpha'\beta_+) <\, \lambda_1 < 1/\eta(1 - \alpha\beta_+),
                \end{equation}
    whence $\eta(1 - \alpha\beta_+) < 1 + \alpha'\beta_+$ or $\beta_+ > (\eta - 1)/(\alpha' + \eta\alpha)$. In addition, taking into account that $\beta_+ < 1$ by (3.11), we have
               \begin{equation}
                        1 > \beta_+ > \frac{\eta - 1}{\alpha' + \eta\alpha}\,,
                        \vspace{-5pt}
               \end{equation}
    from which it follows that
               \vspace{-3pt}
               \begin{equation}
                        \alpha' + \eta\alpha > \eta - 1.
                        \vspace{-1pt}
               \end{equation}

        Now recall that, according to (3.9) and (3.10), $\alpha > 0$ and $\alpha' > 0$. Hence
               \begin{equation}
                         \alpha + \alpha' > 0.
               \end{equation}
        Relation (3.18) is compatible with (3.19) if and only if $\eta = 1$.
        Consequently, the above assumption $\lambda_2 = \eta\lambda_1$ results in $\lambda_2 = \lambda_1$. By inserting this into (2.7) and using notation (3.11), we obtain \vspace{-2pt}
                \begin{equation}
                \lambda_1 = \lambda_2 = \gamma_+\! \equiv \frac{1}{{\sqrt {\smash[b]{1 - \beta_+^2}}}}\,.
                \end{equation}


                                \section{\Large The Lorentz transformations}
    Now substitution of (3.20):

    (i) \;into (1.3) and (3.8), and then

    (ii) into (1.2) and (3.7)

    \noindent results in the Lorentz-like direct and inverse transformations, respectively,
                    \begin{gather}
                        x'(t')  =  \gamma_+(x(t)  -  vt),  \\
                        t' = \gamma_+(t - vx(t)/c_+^2),
                    \end{gather}
                        \vspace{-22pt}
                    \begin{gather}
                        x(t)  =  \gamma_+(x'(t') + vt'),  \\
                        t = \gamma_+(t' + vx'(t')/c_+^2).
                    \end{gather}

    In a combined theoretical and experimental study [8], the speed $c_{em}$ of electromagnetic radiation in a vacuum\footnote{\,It is conventionally assumed that $c_{em} = c$, where $c = 299,792,458$ m/s is the speed of light in a vacuum.} was compared to the limiting speed $c_m$ of massive particles. It was found that $1 - c_m/c_{em} = 1 - c_m/c = 1(12)\times10^{-6}$. Even a more stringent constraint\, $\mid\! 1 - c_m^2/c_{em}^2\!\mid\; < 3 \times 10^{-22}$\, was obtained in [9]. It follows that to a high degree of accuracy $c_m = c$.

    If one assumes now that $c_+ = c_m$, then, accordingly, $c_+ = c$ and thus $\beta_+ = \beta$, where $\beta = v/c$. As a result, (3.20) becomes
                                \vspace{-1pt}
                \begin{equation}
                       \lambda_1 = \lambda_2 = \gamma \equiv \frac{1}{{\sqrt {\smash[b]{1 - \beta^2}}}}\,,
                \end{equation}
    where $\gamma$ is the famous Lorentz factor. Substitution of (4.5) into (4.1)--(4.2) and then into (4.3)--(4.4) leads respectively to the direct and the inverse non-relativistic Lorentz transformations
                \begin{gather}
                        x'(t')  =  \gamma(x(t)  -  vt), \\
                        t' = \gamma(t - vx(t)/c^2),
                \end{gather}
                        \vspace{-23pt}
                \begin{gather}
                        x(t)  =  \gamma(x'(t') + vt'),  \\
                        t = \gamma(t' + vx'(t')/c^2).
                \end{gather}

    Next, if we drop the arguments $t$ in $x(t)$ and $t'$ in $x'(t')$, then (4.6)--(4.7) and (4.8)--4.9) will take the respective form
                \begin{gather}
                        x'  =  \gamma(x - vt),  \\
                        t' = \gamma(t - vx/c^2),
                \end{gather}
                        \vspace{-24pt}
                \begin{gather}
                        x  =  \gamma(x' + vt'),  \\
                        t = \gamma(t' + vx'/c^2)
                \end{gather}
   coinciding with the corresponding relativistic Lorentz transformations [6, pp. 236 and 237, Eqs. (70a) and (70b)].

   It should, however, be kept in mind that the relativistic Lorentz transformations (RLT) on the one hand and Eqs. (4.5)--(4.13) on the other hand deal with different objects:

        1. The RLT are concerned with some \textbf{fixed event}
        considered in the above inertial Cartesian frames of reference S and S', whereas

        2. Relations (4.5)--(4.13) refer to a certain \textbf{material point}
        \textbf{moving} with respect to the same reference frames S and S'.

        3. Although the upper bound\, $c_+$ in  (4.5)--(4.13) is taken equal to the speed $c$\, of light in a vacuum, it has nothing to do with light as an electromagnetic phenomenon which plays the major role in Einstein's special relativity in general and in the relativistic Lorentz transformations in particular.
                                \section{\Large Einstein's theorem of velocity addition}
    Replacing in (2.9)--(2.10) the upper bound \,\!\! $c_+$ by $c = $\, 299,792,458 m/s and dropping the arguments $t$ and $t'$, we have
                \begin{gather}
                        u = \frac{u' + v}{1 + u'v/c^2}\,,\\
                        u' = \frac{u - v}{1 - uv/c^2}.
                \end{gather}
    These formulae obtained from the non-relativistic theorem of velocity composition (2.9)--(2.10) coincide with Einstein's relativistic theorem of velocity addition [2, p. 423].\footnote{\,The remarks made at the end of preceding Section 4 about Eqs.\!\! (4.5)--(4.13) versus the RLT also hold for Eqs. (5.1)--(5.2) versus Einstein's relativistic theorem of velocity addition\,.}

                                \section*{\Large Conclusion}
        In this paper, the theorem of velocity composition (2.9)--(2.10), the Lorentz transformations (4.5)--(4.13) and Einstein's theorem of velocity addition (5.1)--(5.2) are derived, for the first time, within the framework of classical kinematics as applied to the rectilinear compound motion of a material point whose relative and absolute speeds have an upper bound $c_+$. The latter can coincide, in particular, with the speed $c$\, of light in a vacuum.

                 \makeatletter 
                 \renewcommand\@biblabel[1]{#1.} 
                 \makeatother


                    \end{document}